\documentclass[fp]{jpsj3}
\usepackage{txfonts}

\usepackage{color}

\title{Investigating Network Parameters in Neural-Network Quantum States}

\author{Yusuke Nomura$^1$\thanks{yusuke.nomura@riken.jp}}
\inst{$^1$RIKEN Center for Emergent Matter Science, 2-1 Hirosawa, Wako, 351-0198, Japan} %\\

\abst{
Recently, quantum-state representation using artificial neural networks has started to be recognized as a powerful tool.  
However, due to the black-box nature of machine learning, it is difficult to analyze what machine learns or why it is powerful. 
Here, by applying one of the simplest neural networks, the restricted Boltzmann machine (RBM), to the ground-state representation of the one-dimensional (1D) transverse-field Ising (TFI) model, we make an attempt to directly analyze the optimized network parameters. 
In the RBM optimization, a zero-temperature quantum state is mapped onto a finite-temperature classical state of the extended Ising spins that constitute the RBM. 
We find that the quantum phase transition from the ordered phase to the disordered phase in the 1D TFI model by increasing the transverse field is clearly reflected in the behaviors of the optimized RBM parameters and hence in the finite-temperature phase diagram of the classical RBM Ising system. 
The present finding of a correspondence between the neural-network parameters and quantum phases suggests that a careful investigation of the neural-network parameters may provide a new route to extracting nontrivial physical insights from the neural-network wave functions.
}

\begin{document}
\maketitle

\section{Introduction}
\label{sec:intro}

Quantum many-body wave functions are key quantities in understanding quantum many-body phenomena. 
Indeed, good ansatzes for the quantum many-body wave functions such as the Bardeen-Cooper-Schrieffer~\cite{Bardeen_1957} and Laughlin~\cite{Laughlin_1983} wave functions have promoted our understanding of physics. 
Although these wave functions were constructed by human brains, the recent rapid development of numerical methods has offered new routes to construct accurate wave functions. 
In particular, using machine learning techniques for this purpose is quite interesting because the use of ``machine brains'' might shed new light on unsolved problems in physics. 

Such a trend began in 2017 when Carleo and Troyer introduced quantum many-body wave functions constructed from artificial neural networks for quantum spin systems without frustration~\cite{Carleo_2017}.  
Since then, much effort has been made to improve the performance and extend the applicability of the method. 
Now, the neural-network method has been extended, e.g., to the simulations of 
spin systems with geometrical frustration~\cite{Cai_2018,Liang_2018,Choo_2019,Ferrari_2019,Westerhout_2020,Szabo_2020,Nomura_2021_JPCM,Nomura_2021_PRX,Astrakhantsev_2021,M_Li_arXiv}, 
itinerant boson systems~\cite{Saito_2017,Saito_2018}, 
fermion systems~\cite{Cai_2018,Nomura_2017,Luo_2019,Han_2019,Choo_2020,Pfau_2020,Hermann_2020,Yoshioka_2021,Stokes_2020,Inui_2021},
fermion-boson coupled systems~\cite{Nomura_2020}, 
topologically nontrivial quantum states~\cite{Deng_2017,Deng_2017_2,Glasser_2018,Clark_2018,Sirui_2019,Kaubruegger_2018,Huang_2021},
excited states~\cite{Choo_2018,Hendry_2019,Nomura_2020,Nomura_2021_JPCM,Vieijra_2020,Yoshioka_2021},
real-time evolution~\cite{Carleo_2017,Czischek_2018,Schmitt_2020},
open quantum systems~\cite{Nagy_2019,Hartmann_2019,Vincentini_2019,Yoshioka_2019},
and 
finite-temperature properties~\cite{Irikura_2020,Nomura_2021_PRL}.
Through various benchmarks using small system sizes that allow the exact diagonalization and special Hamiltonians for which the quantum Monte Carlo calculations can be performed without the sign problem, the usefulness of the artificial neural networks in quantum many-body problems has started to be recognized. 
When the accuracy is ensured, one can go beyond benchmark calculations and apply the artificial neural networks to analyze unsettled quantum many-body problems. 
Indeed, recently, neural-network wave functions have been applied to investigate the physics of frustrated quantum spin systems~\cite{Nomura_2021_PRX,Astrakhantsev_2021}.

While there is growing numerical evidence that artificial neural networks are useful for accurately approximating quantum states, there is little understanding of what the machines have learned.
This is due to the general problem that the machine learning process is a black box.
In order to extract non-trivial physical insights from the many-body wave functions obtained via machine learning, we need to make machine learning ``white box''.
If such a thing becomes possible, it will provide a new perspective in understanding the quantum many-body problems.

In this paper, as a primitive step, we take one of the simplest neural networks, the restricted Boltzmann machine (RBM), to learn the ground state of the 1D TFI model, whose ground state properties are well understood.  
The RBM is a generative model, which exploits the Boltzmann weights of the Ising spin systems consisting of visible and hidden units.  
When the RBM is applied to approximate the ground state wave function of the 1D TFI model, 
the configurations of physical and visible spins are identified, and the zero-temperature quantum state is mapped onto an ensemble of finite-temperature classical states of the extended Ising spin system (see Sec.~\ref{sec:method} for more details).

Although the RBM wave function is successfully applied to the ground-state representation of quantum spin models~\cite{Carleo_2017,Nomura_2021_JPCM}, the coupling parameters of the optimized RBM network have seldom been investigated. 
Here, by applying the RBM wave function to the 1D TFI model, we directly investigate the property of the RBM parameters. 
We find that a quantum phase transition in the 1D TFI model is indeed reflected in the optimized RBM; the behavior of the tail of the RBM coupling parameters changes qualitatively associated with the quantum phase transition. 
As a result, the finite-temperature phase diagram of the RBM spin system changes qualitatively at the quantum phase transition point. 
The present result suggests the potential importance of analyzing the optimized neural-network parameters themselves. 

This paper is organized as follows. 
Sec.~\ref{sec:method} introduces the RBM wave function and shows that a zero-temperature quantum state is mapped onto a finite-temperature classical state of the extended Ising spins that constitute the RBM. 
In Sec.~\ref{sec:results}, we apply the RBM wave function to approximate the ground state of the 1D TFI model. 
We then show the results of the investigation of the optimized RBM parameters and the finite-temperature phase diagram of the RBM spin system. 
Sec.~\ref{sec:summary} is devoted to the discussion and summary.

\section{Method}
\label{sec:method}

\subsection{Zero-Temperature Quantum State as Finite-Temperature RBM Classical State}
\label{sec:method1}

\begin{figure}[t]
\begin{center}
\includegraphics[width=0.45\textwidth]{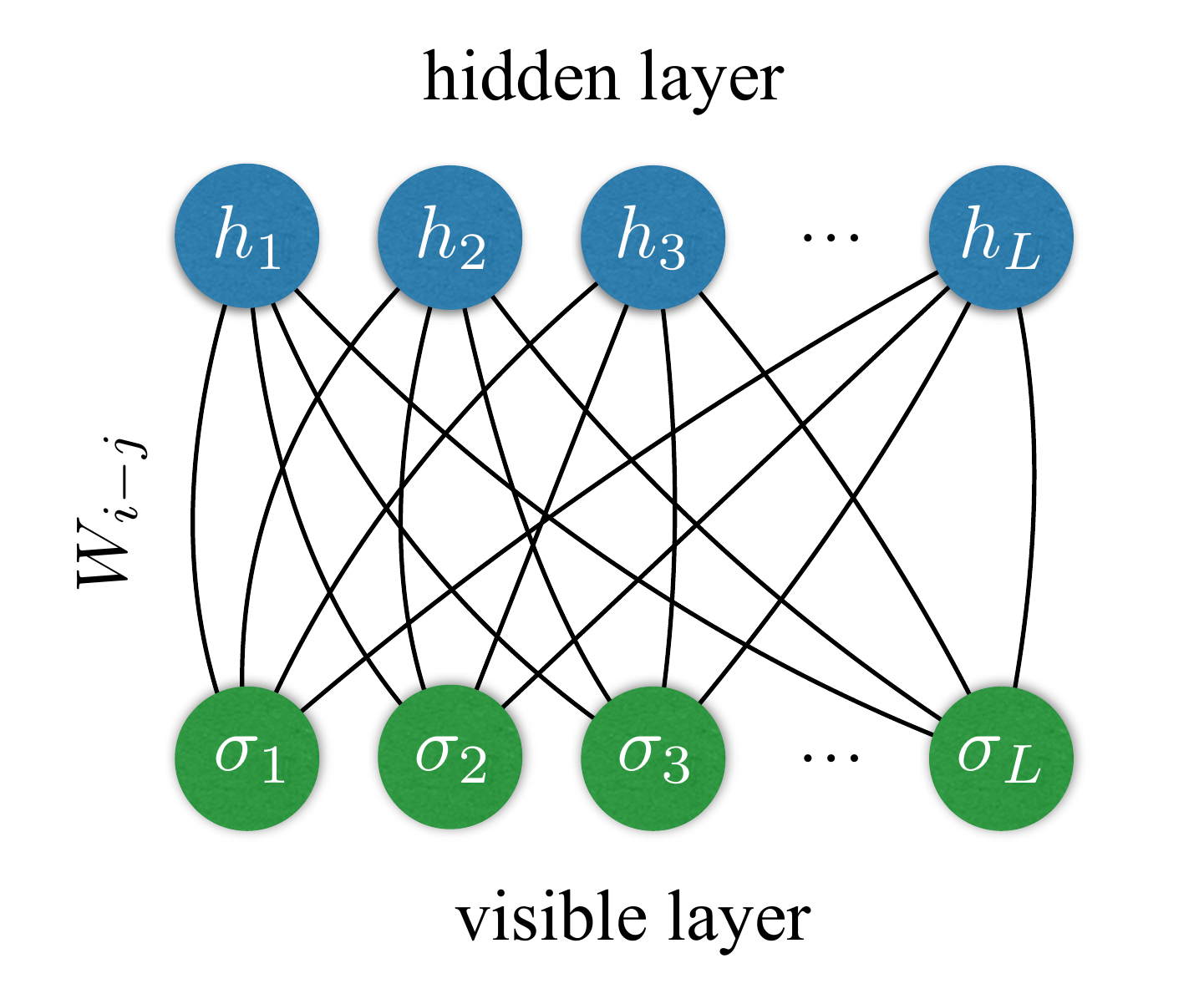}
\caption{
(Color online)
Structure of RBM employed in the present study. 
To represent the wave function of the 1D TFI model, we identify the visible spin configuration with that of spins on the periodic $L$-spin chain ($\varv = \sigma^z$). 
The number of hidden spins is taken to be the same as the physical spins. 
The translational symmetry is imposed in real coupling parameters, and we ignore bias terms. 
Then, the variational parameters are $W_{i-j}$ with $i-j=0, 1, \ldots, L-1$.
The RBM energy and the RBM wave function are given by $E_{\rm RBM}(\sigma^z,h) =   -  \sum_{i,j} W_{i-j}  \sigma^z_i h_j$ [Eq.~(\ref{Eq:RBM_ene})], and $\Psi(\sigma^z)  =   \sum_{ \{ h_j\}}  \exp \Bigl ( -E_{\rm RBM} (\sigma^z, h) \Bigr ) $ [Eq.~(\ref{Eq:RBM_wf})], respectively (see text for more detail).  
}
\label{fig:RBM_structure}
\end{center}
\end{figure}

The RBM consists of visible and hidden layers [see Fig.~\ref{fig:RBM_structure} for the structure of the RBM employed in this paper]. 
The visible and hidden units can be viewed as classical Ising spins ($v_i = \pm1$, $h_j =\pm 1$).
Therefore, hereafter, we will refer to these degrees of freedom as spins.
In the RBM, we consider the Ising spin model for visible and hidden spins, whose energy is given by 
\begin{eqnarray}
E_{\rm RBM}(\varv,h) =   - \sum_i a_i \varv_i -  \sum_{i,j} W_{ij}  \varv_i h_j  - \sum_{j} b_j h_j, 
\end{eqnarray}
where $a_i$ ($b_j$) and $W_{ij}$ are magnetic field for the visible (hidden) spins and Ising-type interactions between visible and hidden spins, respectively. 
$\varv$ and $h$ denote visible and hidden spin configurations, respectively. 
For $S=1/2$ quantum spin models, by identifying the visible and physical spin configurations ($\varv = \sigma^z$), 
the RBM wave function can be defined as (we omit the normalization factor for simplicity)~\cite{Carleo_2017}
\begin{eqnarray}
\label{Eq:RBM_wf}
 \Psi(\sigma^z) \! \! \! \!  &=& \! \! \!  \!   \sum_{ \{ h_j\}}  \exp \Bigl ( -E_{\rm RBM} (\sigma^z, h) \Bigr )  \\ 
 &=& \! \! \! \!  e^{\sum_i a_i \sigma_i^z } \prod_j   2 \cosh \left( b_j + \sum_{i} W_{ij}  \sigma^z_i  \right ).   
\end{eqnarray}
One can see that a quantum state $\Psi(\sigma^z)$ is represented by a thermal state with a temperature $T$ of 1 (i.e., $\beta = 1/T = 1$).

In this study, as a first step, we consider a particular case, where the wave function is positive definite  ($\Psi(\sigma^z)> 0$), which is realized in, e.g., the ground state of the quantum spin systems without frustration.  
In this case, we can take all the variational parameters \{$a_i$, $b_j$, $W_{ij}$\} to be real variables. 
Furthermore, if the Hamiltonian of the quantum spin systems has the translational symmetry and the time-reversal symmetry, we can impose the translational symmetry in $W_{ij}$ parameters and ignore $a_i$ and $b_j$ terms. 
Then, the energy of the classical RBM system is simplified to 
\begin{eqnarray}
\label{Eq:RBM_ene}
  E_{\rm RBM}(\sigma^z,h) =   -  \sum_{i,j} W_{i-j}  \sigma^z_i h_j.  
\end{eqnarray}
In the RBM representation, since the quantum states are mapped onto thermal states of the classical RBM Ising system (quantum-to-classical mapping within the representability of the RBM) as described above, it is of great interest to investigate how the properties of the classical RBM system differ depending on the nature of the quantum states.

\subsection{Analysis of the RBM System Applied to the 1D TFI Model}

In this study, we investigate the 1D TFI model, whose Hamiltonian reads  
\begin{eqnarray}
{\mathcal H}_{\rm TFI} = - J \sum_i \sigma^z_i \sigma^z_{i+1} - \Gamma \sum_i \sigma^x_i   \quad (\Gamma > 0)
\end{eqnarray}
where $J$ and $\Gamma$ are the strength of the Ising-type interaction between the neighboring spins and the transverse field, respectively. 
We consider the periodic boundary condition, and the length of the chain is denoted as $L$ ($\sigma_{L+1} = \sigma_1$). 
Here, we take $J$ as the energy unit ($J=1$).

In the 1D TFI model at zero temperature, as the transverse field $\Gamma$ increases, there exists a quantum phase transition at $\Gamma=1$ from an ordered phase (ferromagnetic state) to a disordered phase, in which the ferromagnetic order in the $z$ direction is quenched due to the quantum fluctuation. 
It is a nontrivial question how the quantum phase transition is reflected in the properties of the RBM Ising system whose couplings are optimized to approximate the ground state of the 1D TFI model.

The analyses are performed as follows: 
First, we optimize the RBM wave function in Eq.~(\ref{Eq:RBM_wf}) with the RBM energy given by Eq.~(\ref{Eq:RBM_ene}) to approximate the ground state wave function of the 1D TFI model. 
Since the ground state is the lowest-energy eigenstate of the Hamiltonian, we optimize the RBM parameters to minimize the total energy (loss function).
For the initial parameters, we put small random numbers. 
To stabilize the optimization, we employ the stochastic reconfiguration (SR) method~\cite{Sorella_2001}; 
the SR method reproduces the imaginary-time Hamiltonian evolution as accurately as possible within the representability of the RBM variational wave function (see Ref.~\citen{Nomura_2017} for further technical details of the optimization). 
For simplicity, in this paper, the number of hidden spins $N_{\rm h}$ is fixed to be the same as the visible (=system) spins ($N_{\rm h} = L$). 
In this case, the variational parameters are $W_{i-j}$ with $i-j=0, 1, \ldots, L-1$ (see Fig.~\ref{fig:RBM_structure}).
Because all the off-diagonal elements of the Hamiltonian are non-positive in this model, the ground-state wave function becomes positive definite ($\Psi_{\rm GS}(\sigma^z)> 0$), which allows us to set $W_{i-j}$ to real numbers. 

When the optimized RBMs are derived, we analyze the finite-temperature properties of the RBM Ising systems, which consist of $2L$ Ising spins 
($L$ visible and $L$ hidden spins).  
The energy of the RBM spin system is given by Eq.~(\ref{Eq:RBM_ene}). 
Although a particular temperature of $T=1$ is used to approximate the ground-state wave function (as we see above, when the hidden spins are traced out at $T=1$, we obtain the ground state wave function), it would be interesting to investigate the properties of the RBM Ising system at different temperatures.   
For the calculation of the thermal average, we use the standard Metropolis Monte Carlo method.

\section{Results}
\label{sec:results}

\begin{figure}[t]
\begin{center}
\includegraphics[width=0.50\textwidth]{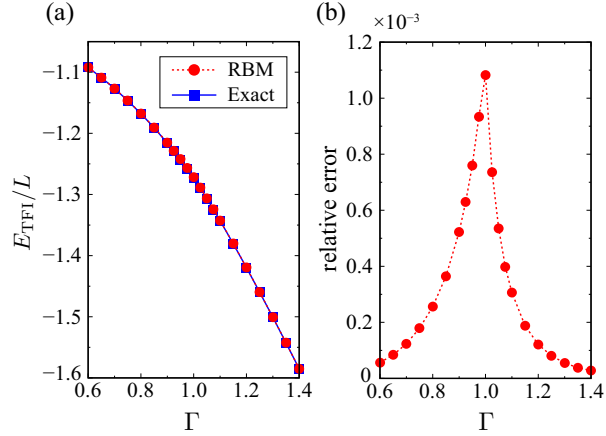}
\caption{
(Color online)
Energy of the optimized RBM wave function as a function of the transverse field $\Gamma$ for the 1D TFI model on the periodic 256-spin chain ($L \! = \! 256$). 
(a) RBM energies (red dots) are compared with exact ground-state energies (blue squares). 
(b) Relative error of the energy $ \bigl |  \bigl ( E_{\rm TFI}({\rm RBM})  - E_{\rm TFI}({\rm exact}) \bigr )    / E_{\rm TFI} ({\rm exact})  \bigr | $.  
}
\label{fig:Energy}
\end{center}
\end{figure}

As is described above, we first optimize the RBM wave function with $L$ hidden spins to approximate the ground-state wave function of the TFI model on the 1D $L$-spin chain with the periodic boundary condition.    
Figure~\ref{fig:Energy}(a) shows the transverse-field $\Gamma$ dependence of the RBM energy at $L=256$.  
We see a good agreement with the exact ground-state energy. 
The relative error of the energy is largest around the critical point $\Gamma =1$; however, the error is at most 0.1 \% [see Fig.~\ref{fig:Energy}(b)]. 
Because the error is small, we can expect that the RBM wave function well captures the ground-state property. 

\begin{figure}[t]
\begin{center}
\includegraphics[width=0.45\textwidth]{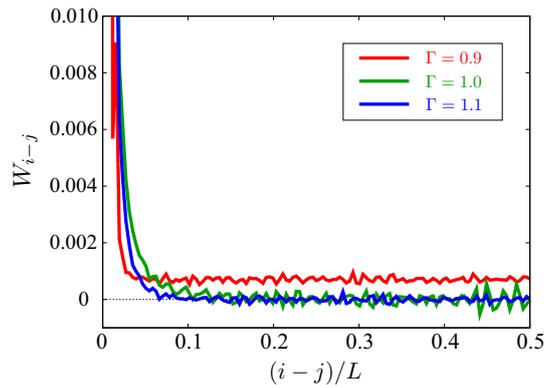}
\caption{
(Color online)
Tail of $W_{i-j}$ parameters of the optimized RBM network to approximate the ground state of the 1D TFI model with $\Gamma=0.9$ (red), $\Gamma=1.0$ (green), and $\Gamma=1.1$ (blue) on the periodic 256-spin chain ($L=256$). 
We define the origin of $i \! -\! j$ such that $| W_0 | $ becomes maximum among $W_{i-j}$ parameters. Note that, due to the translational symmetry, the origin of $i \! - \! j$ is arbitrary (see Fig.~\ref{fig:RBM_structure}).
}
\label{fig:Wtail_raw}
\end{center}
\end{figure}

\begin{figure}[t]
\begin{center}
\includegraphics[width=0.45\textwidth]{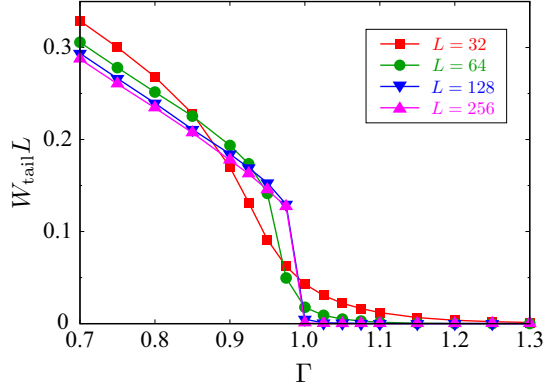}
\caption{
(Color online)
Long-range tail ($W_{\rm tail}$) of the $W_{i-j}$ parameters of the optimized RBM network to approximate the ground state of the 1D TFI model on the periodic $L$-spin chain ($L=32$, 64, 128, and 256). 
$W_{\rm tail}$ is defined as as the average of $W_{i-j}$ parameters in the region $ \frac{L}{2}  \! -\!   \frac{L}{16}  \leq  i \! - \! j  \leq \frac{L}{2}  \! +\! \frac{L}{16} $.
}
\label{fig:Wtail_vs_Gamma}
\end{center}
\end{figure}

Then, it is of great interest to investigate how the optimized $W_{i-j}$ parameters behave. 
Figure~\ref{fig:Wtail_raw} shows the tail of the optimized $W_{i-j}$ parameters for $\Gamma=0.9$, 1.0 and 1.1 at $L=256$ 
[We define the origin of $i \! -\! j$ such that $| W_0 | $ becomes maximum among $W_{i-j}$ parameters. 
Note that, due to the translational symmetry, the origin of $i \! - \! j$ is arbitrary (see Fig.~\ref{fig:RBM_structure})].  
At $\Gamma=0.9$, we find that $W_{i-j}$ saturates at a finite value, and thus, the coupling of the RBM Ising system becomes long-ranged. 
On the other hand, at $\Gamma=1.1$, the tail of $W_{i-j}$ vanishes, and the coupling becomes short-ranged. 
As we see below (Fig.~\ref{fig:Wtail_vs_Gamma}), the behavior of the long-range tail changes at the quantum phase transition point $\Gamma=1$.

To further analyze the behavior of the long-range tail, we define a quantity $W_{\rm tail}$ as the average of $W_{i-j}$ parameters in the region $ \frac{L}{2}  \! -\!   \frac{L}{16}  \leq  i \! - \! j  \leq \frac{L}{2}  \! +\! \frac{L}{16} $.
Figure~\ref{fig:Wtail_vs_Gamma} shows the system-size dependence of $W_{\rm tail}$. 
Whereas $W_{\rm tail}$ shows a gradual crossover in the case of $L=32$, $W_{\rm tail}$ shows a sharp change at the critical point $\Gamma=1$ as the system size increases. 
When the ground state of the TFI model has a ferromagnetic order ($\Gamma<1$), the optimized RBM spin system is characterized by long-range ferromagnetic coupling
[To be more precise, the tail of $W_{i-j}$ parameters scales as $1/L$ (Figure~\ref{fig:Wtail_vs_Gamma} indicates that $W_{\rm tail}L$ converges to a finite value), so it would vanish in the thermodynamic limit. However, the tail part gives a finite contribution to the total energy of the RBM spin system].
On the other hand, when a disordered state is realized in the ground state ($\Gamma>1$), the coupling of the RBM spin system becomes short-ranged. 
We thus find that the ferromagnetic order in the ground state of the TFI model for $\Gamma<1$ is mediated by the long-range coupling of the RBM system and that the quantum phase transition gives a drastic change in the character of the optimized RBM spin system.

We then analyze the finite-temperature property of the RBM spin system (Fig.~\ref{fig:RBM_structure}) consisting of $2L$ classical spins ($L$ visible and $L$ hidden spins).
The energy is given by $E_{\rm RBM}(\sigma^z,h) =   -  \sum_{i,j} W_{i-j}  \sigma^z_i h_j$ [Eq.~(\ref{Eq:RBM_ene}), repeated here for convenience]. 
As we discussed in Sec.~\ref{sec:method1}, Boltzmann weight of the RBM system at a temperature of $T=1$ is employed to represent the wave function $\Psi(\sigma^z) =  \sum_{ \{ h_j\}}  \exp \Bigl ( -E_{\rm RBM} (\sigma^z, h) \Bigr ) $  
[Eq.~(\ref{Eq:RBM_wf}), repeated here for convenience]. 
Here, instead of analyzing the RBM wave function $\Psi(\sigma^z)$, we analyze the property of the RBM spin system itself.

\begin{figure}[t]
\begin{center}
\includegraphics[width=0.5\textwidth]{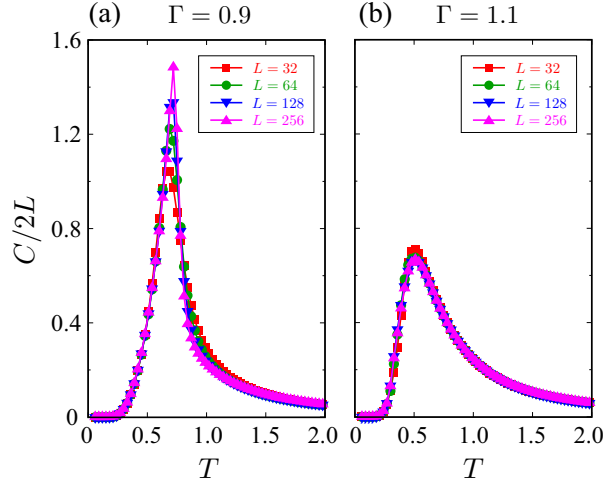}
\caption{
(Color online)
Specific heat $C$ computed from the energy variance as  $C= \frac{1}{T^2}  \bigl ( \langle  {\mathcal H}_{\rm RBM}^2 \rangle -  \langle  {\mathcal H}_{\rm RBM} \rangle^2 \bigr)$ per site of the RBM Ising spin system (Fig.~\ref{fig:RBM_structure}), which is optimized to approximate the ground state of the 1D TFI model on the periodic $L$-spin chain ($L=32$, 64, 128, and 256). 
%The RBM spin system (Fig.~\ref{fig:RBM_structure}) consists of $2L$ Ising spins ($L$ visible and $L$ hidden spins).
The panels (a) and (b) show the results for $\Gamma=0.9$ and $\Gamma=1.1$, respectively. 
}
\label{fig:Specific_Heat}
\end{center}
\end{figure}

Figure~\ref{fig:Specific_Heat} shows the system size dependence of the specific heat $C$ computed from the energy variance as $C = \frac{1}{T^2}  ( \langle  {\mathcal H}_{\rm RBM}^2 \rangle -  \langle  {\mathcal H}_{\rm RBM} \rangle^2)$ per site of the optimized RBM Ising spin system for (a) $\Gamma=0.9$ and (b) $\Gamma=1.1$. 
In the case of $\Gamma=0.9$, since the coupling of the RBM system becomes long-ranged, 
there exists a finite-temperature phase transition from a paramagnetic state to a ferromagnetic state in the thermodynamic limit. 
We see that the temperature of $T=1$, which is used to represent the ground-state wave function of the TFI model, belongs to the paramagnetic region.
On the other hand, when $\Gamma=1.1$, because the coupling is short-ranged, there is no phase transition at finite temperatures, and the specific heat shows a crossover behavior.

\begin{figure}[t]
\begin{center}
\includegraphics[width=0.45\textwidth]{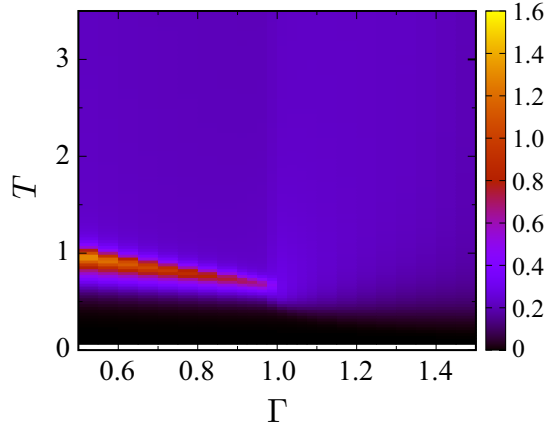}
\caption{
(Color online)
Energy variance $ \langle  {\mathcal H}_{\rm RBM}^2 \rangle -  \langle  {\mathcal H}_{\rm RBM} \rangle^2$ ($= CT^2$) per site of the RBM Ising spin system, which is optimized to approximate the ground state of the 1D TFI model on the periodic 256-spin chain ($L=256$). 
}
\label{fig:Variance_vs_T}
\end{center}
\end{figure}

Figure~\ref{fig:Variance_vs_T} shows the energy variance $\langle {\mathcal H}_{\rm RBM}^2 \rangle -  \langle {\mathcal H}_{\rm RBM} \rangle^2 $ ($=CT^2$) per site of the optimized RBM system as a function $\Gamma$ at $L=256$. 
We clearly see a qualitative change between $\Gamma<1$ and $\Gamma>1$, reflecting the long-ranged and short-ranged couplings in the former and latter cases, respectively. 
Thus, we see that the quantum phase transition at $\Gamma=1$ in the 1D TFI model gives a drastic change in the finite-temperature phase diagram of the classical RBM spin system. 
Conversely, one can detect the quantum phase transition from the change in the property of the classical RBM spin system\cite{note_criticality}.

\section{Discussion and Summary}
\label{sec:summary}

In the present study, by applying the RBM wave function to learn the ground state of the 1D TFI model, we have shown that the quantum phase transition is encoded in the RBM spin system. 
We note that Refs.~\citen{Tanaka_2017,Arai_2018,Miyajima_2021} had already shown that one can detect (quantum) phase transition(s) from the neural-network parameters optimized to relate spin configurations and the corresponding parameters characterizing the spin system, such as temperature and Hamiltonian parameters. 
The current approach is different from those previous studies. 
Here, the output of the neural network is not the parameters of the system but the quantum many-body wave function. 
If the optimization (learning) is successful and the accuracy of the RBM wave function is good, the RBM wave function should change its property at the quantum phase transition. 
However, in the present study, we also investigate the RBM parameters that produce the wave function and find that the RBM parameters themselves change their qualitative property.

This is noteworthy because such qualitative change is not seen in, e.g., the path-integral formalism, another quantum-to-classical mapping than the present approach (we can view the RBM representation of quantum states as a kind of quantum-to-classical mapping within the representability of the RBM, see Sec.~\ref{sec:method1}). 
In the path-integral formalism, the 1D TFI model is mapped onto a classical two-dimensional spin system, whose coupling parameters change gradually as a function of the transverse field (crossover-like behavior). 
The qualitative change at the quantum phase transition in the present case indicates that the mapping using the RBM, which is a more compact mapping using $2L$ classical spins (the path-integral formalism employs $N_{\tau} L$ spins with the number of Suzuki-Trotter slices $N_{\tau}$), indeed achieves a nontrivial mapping through the optimization (learning).

The present finding of the close relationship between the quantum phases and the neural-network parameters is encouraging because not only the wave function (output of the neural network) but also the neural-network parameters themselves have some useful information.  
Therefore, the careful analysis of the neural-network parameters may provide a new route to extracting nontrivial physical insights from the neural-network wave functions.
It is of great interest to also investigate other neural networks and other quantum-spin Hamiltonians and investigate how the neural-network parameters behave.

\begin{acknowledgment}

%\acknowledgment

{\it Acknowledgments.}
We are grateful for the helpful discussions with Nobuyuki Yoshioka. 
This work was supported by Grant-in-Aids for Scientific Research (JSPS KAKENHI) (Grants No. 16H06345, No. 20K14423 and No. 21H01041) and MEXT as “Program for Promoting Researches on the Supercomputer Fugaku” (Basic Science for Emergence and Functionality in Quantum Matter ---Innovative Strongly-Correlated Electron Science by Integration of “Fugaku” and Frontier Experiments---, JPMXP1020200104). 

\end{acknowledgment}

\bibliographystyle{jpsj}
\bibliography{main}

\begin{thebibliography}{10}
\bibitem{note_criticality}
Because the relative error of energy of the optimized RBM wave function for the
  1D TFI model has a sharp peak at $\Gamma=1$ [see Fig.~\ref{fig:Energy}(b)],
  the derivative of the energy of the RBM wave function with respect to the
  $\Gamma$ parameters has a small discontinuity at $\Gamma=1$. Accordingly, the
  property of the optimized RBM spin system changes discontinuously at
  $\Gamma=1$ (thus, the $W_{\rm tail}$ value has a jump at $\Gamma=1$, see
  Fig.~\ref{fig:Wtail_vs_Gamma}). Therefore, although one can detect the
  quantum phase transition using the RBM, the critical behavior of the 1D TFI
  model at $\Gamma=1$ cannot be captured in the current setup of the RBM.
\end{thebibliography}

%\begin{thebibliography}{9}
%\bibitem{jpsj} The abbreviation for JPSJ must be ``J. Phys. Soc. Jpn." 
%\end{thebibliography}

\end{document}